\documentclass{WileyMSP-template}

\usepackage[numbers]{natbib}
\usepackage{graphicx}
\usepackage{amssymb}
\usepackage{lineno}
\usepackage{amsthm}
\usepackage[normalem]{ulem}
\usepackage{xcolor}
\usepackage{balance}
\usepackage{hyperref}

\begin{document}

\begin{figure}[t!]
    \centering
    \includegraphics[width=0.9\textwidth]{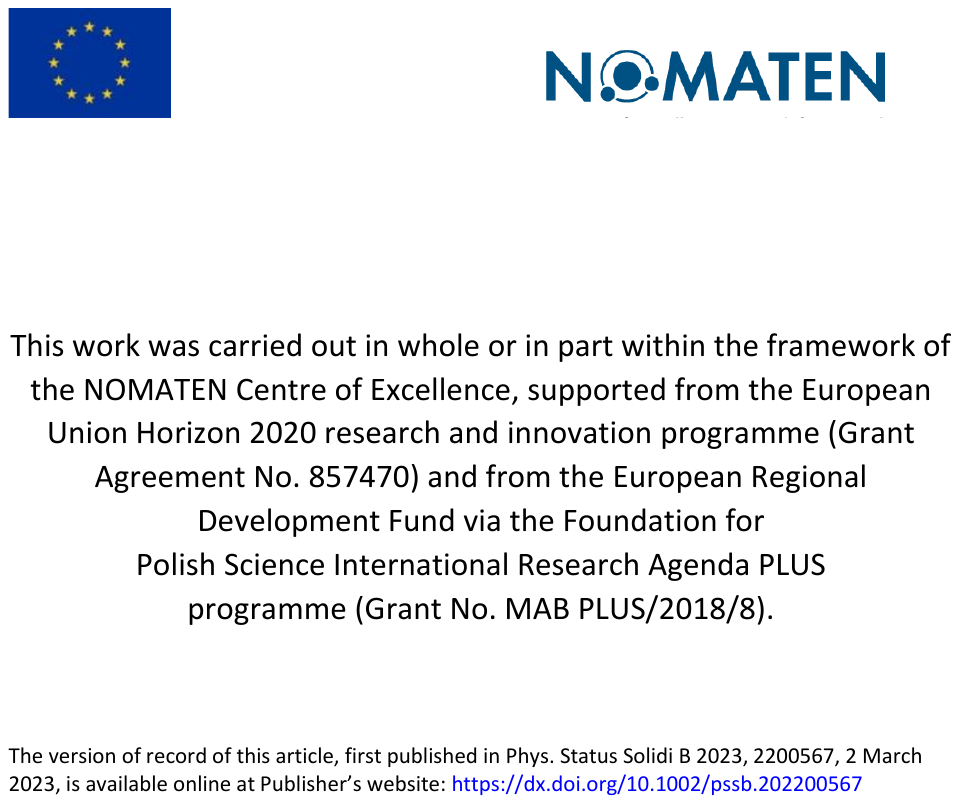}
\end{figure}

\pagestyle{fancy}
\rhead{\includegraphics[width=2.5cm]{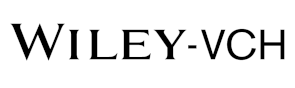}}

\title{Dynamical pathways for the interaction of O$_2$, H$_2$O, CH$_4$, 
and CO$_2$ with $\alpha$-alumina surfaces: Density functional tight--binding  calculations}

\maketitle


\author{F. Javier Dom\'inguez--Guti\'errez*,}
\author{Amil Aligayev,}
\author{Wenyi Huo,}
\author{Muralidhar Chourashiya,}
\author{Qinqin Xu,}
\author{and Stefanos Papanikolaou}



\begin{affiliations}
F. J. Dom\'inguez--Guti\'errez, A. Aligayev, 
Q. Q. Xu, M. Chourashiya, W. Y. Huo, S. Papanikolaou\\
NOMATEN Centre of Excellence, National Centre for Nuclear Research, ul. A.
Soltana 7, Otwock, Poland\\
Email Address: javier.dominguez@ncbj.gov.pl

F. J. Dom\'inguez--Guti\'errez\\
Institute for Advanced Computational Science, Stony Brook University, Stony Brook, NY 11749, USA

A. Aligayev\\
University of Science and Technology of China, Hefei, 230026, China

M. Chourashiya\\
Technion, Israel Institute of Technology, Haifa, 32000, Israel\\
Guangdong Technion Israel Institute of Technology, Shantou, 515063, China

W. Y. Huo\\
College of Mechanical and Electrical Engineering, Nanjing Forestry University, Nanjing, 210037, China

\end{affiliations}



\begin{abstract}

In this study, we investigated the physisorption mechanisms of O$_2$, 
H$_2$O, CH$_4$, and CO$_2$ molecules on alumina and their effect on
electronic properties. We employed quantum-classical molecular dynamics 
simulations and the self-consistent-charge density-functional 
tight-binding (SCC-DFTB) approach to dynamically model these mechanisms. 
Our results revealed the binding pathways of O, H, and C atoms in the various 
molecules to Al and O atoms at the top atomic layers of the $\alpha-$alumina 
surface. We examined several adsorption sites and molecular orientations 
relative to Al-terminated and Ox-terminated alumina surfaces and found
that the most stable physisorbed state on the Al-terminated surface is
located above the 
Al atom, while the Ox-terminated state is found above the oxygen,
resulting in enhanced optical adsorbance. 
The dissociation of CH$_4$ into CH$_2$+H$_2$ after 
interaction with the surface resulted in hydrogen production, but 
with low adsorbate rates. While, O$_2$ molecules primarily bond to
the Al atoms, leading to the highest adsorbance rate among the
other molecules.
Our findings provide important insights into the physisorption
mechanisms of molecules on alumina and their impact on
electronic properties.

\end{abstract}


\section{Introduction}
\label{sec:intro}

$\alpha$-Alumina, i.e., corundum, is one of the most widely 
used ceramic materials due to its excellent properties, such 
as high chemical stability, high thermal resistance, and
high electrical resistance, and is considered to be an excellent adsorbent, 
with primary examples being radiation, high temperature
and corrosion 
\cite{thomas1993physical,evans1995review,lin2004crystal,guo2009one,terrani2013protection,zaborowska2021absolute}.
The interface reactions of oxide/molecules 
dominate in many fields, such as lubricants 
\cite{blanck2020adhesion,rey2021transferable}, 
electronic industries \cite{xu2018theoretical}, 
automobile engines 
\cite{munitz1979interface,shorowordi2003microstructure} 
and other transportation systems 
\cite{huo2022a,su2015enhancing}. 
This material is utilized as an adsorbent, catalyst, 
and catalyst support in several environmentally 
\textcolor{blue}{environmentally important}.
. Ferri \textit{et al.} studied CO adsorption
on catalytic solid/liquid interfaces by performing 
ATR-IR spectroscopy \cite{ferri2001pt}. Mason 
\textit{et al.} reported the adsorption energy of 
Pb(II) onto the \{0001\} surface of $\alpha$-Al$_{2}$O$_{3}$ 
and $\alpha$-Fe$_{2}$O$_{3}$ based on DFT 
calculations \cite{mason2009pb}.
Moreover, alumina is also used as a desiccant, which is 
a substance that absorbs moisture, and in water treatment 
as a filter medium. 
Experimental exploration of aluminum--assisted 
water split reactions has shown that the formation of 
an alumina film inhibits the direct reaction of Al 
with H$_2$O. However, when Al is alloyed with other 
metals (such as gallium, etc.) or prepared 
aluminum--based composites by ball milling
\textcolor{blue}{modify} the
Al--water system for hydrogen production. 
Such technology for onboard hydrogen production is 
of great interest in a hydrogen economy 
\cite{DUPIANO20114781}. 
Finally, optical adsorption on alumina surfaces is a 
process where light is absorbed 
\textcolor{blue}{resulting in} to a reduction in the 
light intensity as it passes through. This absorption provides
useful information about the electronic and optical properties 
of the alumina surface, including its bandgap, surface state, 
and surface morphology where different molecules can 
affect these properties \cite{TongY,QiD,TYujin}.

\textcolor{blue}{Atomistic simulations are now used as a guide for 
experimental adsorption measurements, which are complicated to
carry out for alumina materials due to the complex crystal
structure and surface defects.} 
For this reason, atomistic simulations are important in the
investigation of physisorption processes, making
advantages in understanding the underlying physical
and chemical mechanisms, where alumina surfaces can
be manipulated by the adsorption of different molecules
such as H$_2$O, O$_2$, CO, and CH$_4$ \cite{cabrejas,SHI2020145162,FGang2018}.
Although computations by the density functional density (DFT) 
method can provide accurate results, for large systems 
with open boundaries the computational demands may become 
prohibitively 
expensive and may require the use of specialized 
high-performance computing resources \cite{YU201533}.
Additionally, as the number of atom types in a system 
increases, the number of electrons also increases, and the 
approximations made in DFT calculations may become even 
more computationally expensive.

In several cases, computationally economical 
simulations based on empirical or fitted interatomic 
potentials, such as the ReaxFF method \cite{reaxff} 
can be applied to compute the mechanical properties 
of alumina surfaces in good agreement with 
experimental data \cite{LUU2022128342}. 
ReaxFF is based on a bond order formalism to enable the 
description of the bond breaking processes; however, 
charge transfer processes due to oxygen vacancy 
migration require a semi--classical approximation 
for modeling chemical bonding \cite{wang2022small} in more fundamental 
processes. 
Besides, the electronic structure of the system needs 
explicit information 
\cite {wang2022acb,wang2022cej} for electron–electron interactions, 
most notably density functional theory (DFT), as 
a standard software in computational materials science 
with high computational resource demands.
\par

The previous discussion motivates us to utilize the 
self--consistent--charge density--functional tight--binding (SCC--DFTB) 
approach
\cite{DFTBplus,B105782K,PhysRevB.63.085108}, which requires
minimal computational resources 
through DFT packages and provides a connection between
classical approaches and electronic structure 
theory. 
This approach uses a minimal set of electronic states of the
surface that are represented by a set of linear equations 
describing the interactions between the localized orbitals. 
These equations take the form of a Kohn--Sham
Hamiltonian matrix, which is then solved to obtain the 
electronic eigenstates and eigenvalues of the system.
This method is currently a versatile tool for 
investigations in chemistry and materials science 
\cite{doi:10.1021/jp404326d} due to the possibility of 
calculations of large systems and performing simulations 
for longer time scales than DFT 
for extensive studies on dynamical properties in DFTB.
This method has several features that are helpful to 
understand the mechanisms of the molecular adsorption process 
on alumina surfaces, giving us the opportunity to model the
dissociation and formation of \textcolor{blue}{molecules} 
due to the interactions 
of X-atoms with Al and O atoms at the top layer of the surface. 
This type of modeling can be very expensive computationally 
for DFT software requiring long wall times and HPC resources.

In this work, we investigate the interaction at interfaces
of $\alpha$--Al$_{2}$O$_{3}$ with several molecules
based on systematic calculations by DFTB, followed by 
performing quantum-classical molecular dynamics 
simulations of molecular adsorption at room temperature.
This work will help understand the surface interaction
of alumina with water, oxygen, carbon monoxide, and methane that is
essential to design, optimize, and control their 
applications in hydrogen production as well as a 
variety of sensor designs.
The manuscript is organized as follows: In Section 
\ref{sec:meth}, we describe the computational methodology for 
carrying out binding energy calculations by considering 
different physisorption pathways, optical adsorption, and dynamic 
mechanism of molecular adsorption. In Section \ref{sec:results}, 
we present the results for the calculated equilibrium molecule 
surface distance, adsorption rates, and adsorbance with
the corresponding effect on the density of 
defects of the alumina surface. Finally, in Section 
\ref{sec:conl}, concluding remarks are summarized.
\section{Methods}
\label{sec:meth}

The SCC-DFTB method is an approximation to traditional Density Functional 
Theory (DFT) that takes into account valence electron interactions during
dynamics. It involves solving Kohn--Sham equations to obtain total
valence electronic densities and energies for each atom using a Hamiltonian
functional based on a two-center approximation and optimized 
pseudo--atomic orbitals as basis functions. 
Slater--Koster parameter files containing tabulated Hamiltonian matrix
elements, overlap integrals \cite{DFTBplus}, and repulsive splines
fitted to DFT dissociation curves are read into the computer memory
only once at the start of the simulation 
\cite{KOSKINEN2009237}. 
Thus, the total energy of the system is expressed as
\begin{equation}
    E^{\rm DFTB} = E_{\rm band}+E_{\rm rep}+E_{\rm SCC},
\end{equation}
with the band structure energy, $E_{\rm band}$, defined
from the summation of the orbital energies $\epsilon_i$
over all occupied orbitals $\Psi_i$; 
the repulsive energy $E_{\rm rep}$ for the core--core
interactions related to the exchange--correlation energy 
and other contributions in the form of a 
set of distance--dependent pairwise terms;
and an SCC contribution, $E_{\rm SCC}$, as the contributions
given by charge--charge interactions in the system.

In this study, we utilize the SKF pair potentials set for materials 
science simulations (MATSCI) that have been applied to study chemical 
reactions on gibbsite surfaces
\cite{https://doi.org/10.1002/zaac.200500051}, monodentate, bidentate and tridentate
adsorption of the acids on all possible adsorption sites on the alpha--alumina 
surfaces considering different surface coverages \cite{LUSCHTINETZ20081347}, 
and stability, electronic, and mechanical properties of imogolite nanotubes 
\cite{Luciana_ACS}.  
The electronic energies are calculated as a sum over the occupied 
KS single-particle energies and the sum over diatomic repulsive 
energy contributions. 
SCC corrections, as implemented in the DFTB$+$ code ver. 22.2
\cite{DFTBplus}, are included in the total energy via an 
iteration procedure that converges to a new electron density
at every time step during the simulation, where the 
convergence is improved by using an electronic temperature of 
1000 K.

\subsection{Binding energies}

Corundum ($\alpha$--alumina) has a trigonal structure with 
oxygen ions arranged in a hexagonal close packing (HCP). 
Al atoms occupy two--thirds of the octahedral vacancies in
the oxygen sublattice at alternating positions above and 
below the center of these sites. 
The unit cell for alumina is defined at the basal crystal
orientation \{0001\}, which is a crystal plane perpendicular
to the crystal's axis of symmetry and lying in its base. 
With 60 atoms - 24 Al and 36 O - the cell has hexagonal 
symmetry with parameters $a = 0.472$ nm, $c = 1.299$ nm,
$\alpha = \beta = 90^o$, and $\gamma = 120^o$. 
The Al--O bond lengths, 0.0185 nm and 0.0194 nm, were 
optimized using DFTB$+$ with the conjugate gradient 
method (Fig. \ref{fig:fig1}.) and are in good agreement
with experimental data
\cite{LUSCHTINETZ20081347} ($a=0.4763$ nm, $c=1.3$ nm).
In addition, the geometries of the isolated O$_2$, H$_2$O, 
CH$_4$, and CO$_2$ are optimized with an energy convergence
tolerance of 10$^{-6}$ eV. 

\textcolor{blue}{
The total energies, $E(z)$, of the molecule--alumina system with a 
separation $z$ between the top atomic layer and the center of 
mass of the molecule are varied above the surface in a range 
of 0.5 to 7 \AA{}, which defines the computation of the 
adsorption potential as a function of the distance separation.
The total energy is then computed as:}
\begin{equation}
    E(z) = E_{\rm Tot} - \left( E_{\rm Surface} + E_{\rm Molecule}
    \right),
\end{equation}
where $E_{\rm Surface}$ is the total energy of the 
alumina surface; $E_{\rm Molecule}$ is the total 
energy of the isolated molecule: O$_2$, H$_2$O, CH$_4$, 
and CO$_2$; and $E_{\rm Tot}$ is the energy of the 
interacting system at every $z$--distance. 
Thus, the binding energy is defined as 
$E_b = E(z_{\rm min})$ with $z_{\rm min}$ as the 
equilibrium molecule--surface distance. 
Total energy calculations are performed for the 
molecule--alumina system, varying the distance between 
the surface and the center of mass of the molecules 
along the $z$--axis. 
We consider 8 different adsorption sites based on the
hexagonal arrangement of aluminum atoms on the 
Al--terminated surface and the BCC configurations of oxygen
atoms on the Ox-terminated surface. 
The molecular symmetry plane determines the perpendicular
and parallel orientations with respect to the surface plane
in the calculations. The repulsive potential is cut off at
a distance below the second nearest-neighbor interaction
region for numerical stability. 
However, this approximation may not always provide 
satisfactory dissociation curves. 
The SCC-DFTB framework addresses this limitation by 
shifting the repulsive energy functions downward.
\subsection{Optical absorption spectra}
\label{subsec:OAS}

The optical absorption is investigated within the DFTB framework as 
an electronic dynamic process in response to an external electric
field \cite{C8CP04625E,B926051J}. 
The conventional adiabatic approximation gives the time evolution 
of the electron density matrix by time integration of the 
Liouville-von Newmann equation expressed as
\begin{equation}
    i \hbar \frac{\partial \hat \rho}{\partial t} = 
    S^{-1}\hat H \hat \rho - \hat \rho S^{-1},
\end{equation}
where $\hat \rho$ is the single electron density matrix, 
$\hat S$ is the overlap matrix, and $\hat H$ is the system Hamiltonian 
that includes the external electric field as 
$\hat H = \hat H_0 + E_0 \delta (t-t_0) \hat e$ with 
$E_0$, the \textcolor{blue}{magnitude} of the electric field, 
and $\hat e$, its direction. 
Under the framework of linear response, the adsorbance $I(\omega)$ is
calculated as the imaginary part of the Fourier transform of the
induced dipole moment caused by an external field. 
In this study, the external field strength was set to 
$E_0 = 0.001$ V/\AA{}. The induced dipole moment was evaluated over
a $200$ fs time period using a time step of $\Delta t$ = 0.01 fs. 
The Fourier transform was performed with an exponential damping
function (using a 5 fs damping constant) to eliminate noise.

\subsection{Quantum--Classical Molecular dynamics}
\label{subsec:QCMD}

An $\alpha$-alumina surface with a $3\times2\times 1$ supercell
is modeled 
by applying periodic boundary conditions in the x-y direction and 
a 5 nm vacuum above the surface in the z direction, as shown in Fig. 
\ref{fig:fig2}a).
The optimized alumina surface is equilibrated to 300 K (room temperature)
using a Nose-Hoover thermostat. To model the dynamics of molecular
adsorption on the alumina surface, \textcolor{blue}{different molecules
are deposited on a 1 nm$^2$ target area} with a random distribution 
at the center of the surface using the velocity
Verlet algorithm. The time step of 0.25 fs is chosen, and different
molecules are emitted at an impact energy of 1 eV, which is chosen 
respect to the energy bond of the molecules, with 500 independent
trajectories, using the same target. Each molecule is emitted
vertically with a randomly assigned orientation and
an initial distance of 0.6 nm above the surface for
a simulation time of 350 fs. The calculations are performed using
embarrassing parallelization on 160-240 cores of a computer cluster 
with a typical wall time of $\sim$20 min/simulation.

This approach was used in our previous works
to investigate the hydrogenation mechanisms of fullerene 
cages \cite{DOMINGUEZGUTIERREZ2018189} and 
to 
study the electronic properties of the 2D nanomaterial borophene, 
proposed for a boron-based hydrogen detector \cite{C7TC00976C}, 
showing excellent agreement with first principles DFT 
calculations. 
SCC-DFTB, as implemented in the publicly available DFTB$+$ code
version 22.1, is used throughout this work for all geometry 
optimizations and QCMD simulations.
\section{Results and Discussion}
\label{sec:results}

\subsection{Physisorption pathways}
Adsorption calculations were performed on isolated O$_2$, 
H$_2$O, CH$_4$, and CO$_2$ molecules adsorbed on a \{0001\} 
alumina surface. Two cases were considered: an Al--terminated
surface, which is non-dipolar, and an oxygen-terminated
surface, which is dipolar with half of the oxygen atoms removed
from the bottom of the cell. 
The Al--terminated surface is known to be the most stable of
the two, as reported in the literature  
\cite{https://doi.org/10.1111/j.1151-2916.1999.tb02225.x}.
During the computations, a 50 \AA{} vacuum section was
added above the sample to avoid boundary conditions effects 
on the calculations and periodic boundary conditions were
set in the $x$-$y$ directions to simulate a semi-infinite 
surface. A $4\times4\times1$ Monkhorst--Pack set was used for
the $k$-point sampling during all calculations. 
The inter--planar distance of the Al--terminated surface in the
bulk is 2.73 \AA{}, while the top layers have a distance of
1.8 \AA{} due to the vacuum section, which reduces the distance
by $\sim 45 \%$ in agreement with experimental and theoretical data \cite{SongT,https://doi.org/10.1111/j.1151-2916.1999.tb02225.x}. 

Fig. \ref{fig:fig2}a) shows the adsorbate sites considered in this
work for the Al-terminated surface: i) above the top Al atom 
(labeled Al-1); ii) at the second Al layer with an Al atom and 
three oxygen atoms as the nearest neighbors (labeled Al-2); iii) 
the first oxygen layer below the top Al atom, with an oxygen atom
next to an Al atom (labeled O-1); and iv) the void between the
oxygen atoms at the first O layer (labeled O-2).
For the Ox-terminated surface, the optimization process 
does not affect the geometry of the oxygen atoms layered 
at the stab top. 
Here, the Ox--terminated surface is optimized with
an oxygen interlayer distance of 2.176 \AA{} between
the top layers and 2.187 \AA{} between the layers in 
the bulk, in fairly good agreement with DFT calculations, 
reported in the literature 
\cite{https://doi.org/10.1111/j.1151-2916.1999.tb02225.x}.
In Fig. \ref{fig:fig2}b), the adsorbate sites for the Ox--terminated
surface are presented: i) at the top layer, the place above an
oxygen atom defines the O-1 site; ii) the space between the
triangle formed by three oxygen atoms is considered the O-2 site;
iii) due to the symmetry of the alumina surface, the first Al
layer defines the Al-1 site above one Al atom; and iv) the
second Al layer is taken into account by defining Al-2 as the
vacancy located at the middle of the triangle formed by the top
layer's three oxygen atoms and the second layer's six Al atoms.
In Fig. \ref{fig:fig2}c), we present the density of states of the 
supercell sample.

In Fig. \ref{fig:fig3}, we present the energy ($E(z)$) of O$_2$ 
in a), H$_2$O in b), CH$_4$ in c), and CO$_2$ in d) molecules at 
different adsorbate sites on an Al--terminated surface. The 
orientation of the molecules relative to the surface plane is 
considered, both parallel and perpendicular. For O$_2$, the minimum 
contribution is at the adsorbate O--2, where three oxygen atoms 
repulse the molecule. When oriented perpendicular to the surface, the 
O$_2$ molecule can be attracted to the Al--1 site and repelled by the 
Al--2 site. For H$_2$O, the Al--1 and O--1 sites attract the molecule 
due to the vacancy formed by three oxygen and one Al atoms, and the H 
atoms are mainly bonded to O atoms from the surface. For CH$_4$, the 
Al-1 site attracts the molecule due to the interaction of hydrogen 
atoms with the Al atom, but otherwise, the molecule is mostly repelled 
by the surface. 
The results provide insight into the interaction between alumina and 
these molecules, which is important for understanding chemical 
processes such as CO$_2$-CH$_4$ reforming and carbon deposition.
Finally, for CO$_2$, the molecule is adsorbed by the surface when 
oriented perpendicularly at the Al-1 and O-2 sites, but the interaction 
with the C atom and Al is highly repulsive. 

The energy ($E(z)$) of O$_2$, H$_2$O, CH$_4$, and CO$_2$ molecules at 
different adsorbate sites on the Ox--terminated alumina surface is 
presented in Fig. \ref{fig:fig4}. The oxygen molecule can be attracted 
to the O--2 adsorbate site, where an Al atom can adsorb the oxygen 
atom. However, the layer of oxygen atoms can repel the oxygen molecule 
at distances above 0.5 \AA{} regardless of orientation. The stability 
of the O$_2$ molecule is observed at distances above 1.5 \AA{} where 
the Ox--terminated surface has no impact.
Under ultra-high vacuum conditions, the clean \{0001\} surface 
of alpha-alumina is terminated by a single layer of aluminum atoms.
However, in 
the presence of water, the surface is hydroxylated, making the Ox--
terminated surface of interest due to OH bonding. Water molecules are 
observed to prefer a perpendicular orientation at the O--1 and Al--1 
adsorbate sites with the majority of oxygen atoms, as shown in Fig. 
\ref{fig:fig4}b. The Ox--terminated surface does not interact with the 
water molecule at distances above 2 \AA{}, while the Al--terminated 
surface has an interaction range of up to 5.5 \AA{}. The hydrogen 
atoms of the water molecule tend to interact and bond with the oxygen 
atoms at Al-1 and O-2 for the perpendicular orientation, but the 
molecule is mainly repelled by the alumina surface regardless of atom 
termination.
For the Ox--terminated surface, the perpendicularly oriented CO$_2$ 
molecule is attracted to the site between two O atoms in the surface 
and the Al atom at the O-1 site, as shown in Fig. \ref{fig:fig4}d. The 
CO$_2$ molecule is not affected by the Ox--terminated surface for 
distances above 3 \AA{}, which is larger than the ranges for O$_2$, 
H$_2$O, and CH$_4$ molecules.

\subsection{Surface electronic structure}
Table \ref{tab:tab1} reports the binding energy ($E_b$), length ($z_{\rm min}$), 
molecular orientation, and adsorbate site to provide more information about the 
adsorption of molecules by the alumina surface. We also present surface 
energies, computed as $\gamma_{\rm phys} = E_b/A$, where $A$ is the surface 
area \cite{https://doi.org/10.1111/j.1151-2916.1999.tb02225.x}. This value is a combination of the cohesive energy of surface atoms and 
the energy required to create a new surface by breaking bonds and is used to 
determine the sample's equilibrium morphology.
The information in the table is used to energy-optimize the alumina surface with 
different molecules, as shown in Fig. \ref{fig:fig6a} for the Al-terminated 
surface, which is of interest for experiments. The Ox-terminated surface is 
found to be less stable than the Al-terminated surface due to the removal of the 
dipolar nature of the O-terminated plane.

We then performed an optimization process for the Al-terminated 
surface with different molecules. As expected, the most stable point 
for the O$_2$ molecule (O$_1$--O$_2$) was above the Al atom (Al-1 
site) with an internuclear distance of 0.189 nm to the O$_2$ atom, 
which decreased by $\sim 3\%$ compared to the isolated molecule, as 
shown in Fig. \ref{fig:fig6a}a. For the H$_2$O molecule, the distance 
between the O${\rm water}$ and the Al atom at the top of the surface 
was 0.317 nm, as shown in Fig. \ref{fig:fig6a}b. This location of the 
H$_2$O molecules on the Al-terminated surface is related to one of the 
biggest challenges in the direct reaction between water and aluminum, 
where the oxide layer on the aluminum surface prevents penetration 
into the core.

For the CH$_4$ molecule, the bonding between the H atoms and the Al 
atom of the Al-terminated surface was the minimum compared to the 
other molecules, as shown in Fig. \ref{fig:fig6a}c. The bond length of 
the H atoms was not affected by physisorption, with an internuclear 
distance between the most stable adsorbate site (Al-1) and the C atom 
of 0.250 nm. The physisorption of CO$_2$ by the Al-terminated surface 
is depicted in Fig. \ref{fig:fig6a}d, where O$_{\rm CO2}$ is bound to 
the Al atom with an internuclear distance of 0.19 nm. The bond length 
between the C and O atoms was affected by the adsorption of the 
molecule, explaining the observed dissociation of CO$_2$ in 
experiments, where adsorbed O atoms produced CO molecules (CO$_2$ = CO 
$+$ O) \cite{BAHARI2022} with oxygen atoms bound to Al atoms.

The size of the band gap is a crucial parameter in dielectric 
materials, as it affects thermal carrier tunneling. $\alpha$-Alumina 
is considered a good dielectric material with a sufficiently large band gap, 
as demonstrated by the total and partial density of states results 
shown in Fig. \ref{fig:fig6a}. These results are in good agreement 
with DFT calculations \cite{https://doi.org/10.1002/zaac.200500051}.
The adsorption of molecules can greatly alter the density of states (DOS). 
The pristine alumina peak at $+15$ eV shifts to $+1$ eV for 
all cases due to adsorption, which results in a new gap in 
the system as determined by the DFTB approach.

The Mulliken charge of an atom is defined as the sum of the orbital 
populations weighted by their corresponding atomic orbital (AO) 
coefficients. In this study, the DFTB method is used to compute the 
Mulliken charge of individual atoms in the system, as seen in Fig. 
\ref{fig:fig7}a). The histogram of the Mulliken charge of each atoms 
in the system defined by the Al-terminated surface with each 
adsorbed molecule at their binding energy distance. 
It is noted that the oxygen atoms of the alumina surface are highly 
affected by the adsorption of the molecules, with CH$_4$ having the 
greatest effect and CO$_2$ having the least effect on the charge 
distribution of the surface.
The \textcolor{blue}{adsorption} properties of $\alpha$-alumina
can be determined using 
spectroscopic techniques such as UV--visible spectroscopy or infrared 
spectroscopy. These properties can be utilized in the development of 
sensors that respond to specific wavelengths of light and to promote 
chemical reactions when exposed to light. In Fig. \ref{fig:fig7}b, the 
optical adsorption response of alumina with different molecules on its 
surface is presented. It is observed that alumina increases its 
adsorption in the range of 400 to 700 nm, and the presence of O$_2$ 
and CH$_4$ molecules further enhances its adsorption in the visible 
range.
Results for H$_2$O with adsorbance spectra is in a
\textcolor{blue}{qualitatively good agreement} with experimental
data \cite{TYujin} where 
the observed peaks are presumably related to the surface instabilities 
due to the presence of water molecules.


\subsection{Dynamical molecular adsorption}
\label{subsec:collisions}

We compute the probability of reflection and adsorption of 
O$_2$, H$_2$O, CH$_4$, and CO$_2$ molecule emissions on the 
Al--terminated alumina surface at 1 eV of kinetic energy 
to investigate the adsorption and dissociation mechanism 
dynamically at room temperature, using the approach described in
Section 2.3. If the kinetic energy is too low, molecules 
can be only adsorbed with a low probability of dissociation.
The probabilities shown in Fig. \ref{fig:figRA} 
are calculated as the N$_{A,R}$/N$_{\rm Tot}$ ratio, where 
N$_{\rm Tot}$ is the total number of the incident molecules, 
in our case 500 and N$_{A,R}$ is the number of molecules 
which are adsorbed or reflected to the surface. 
Reflected events are determined by measuring a final distance
of 1 nm between the Al atoms at the top layer of the surface
and the center of mass of the molecules.
We notice that O$_2$ molecules are mainly adsorbed by 
the alumina surface and CH$_4$ molecules are less 
adsorbed, in good agreement with the physisorption calculations 
presented in Fig. \ref{fig:fig3}.

In Fig. \ref{fig:islands}, we present the final positions of 
adsorption events for different molecular projectiles. For oxygen 
molecules, we observed a low dissociation probability of 2\% as they 
mainly bond to Al atoms, forming a hexagonal pattern on the surface. 
In contrast, water molecules have a higher dissociation probability
of 40\%, with some of them dissociating into OH$+$H  and producing
H atoms upon interaction with the surface. 
Some water molecules are also both adsorbed and dissociated, leading 
to hydrogen atoms bonding to the oxygen atoms in the second top layer 
of the alumina surface, and a few oxygen atoms penetrating the surface 
and finding a final position next to an oxygen atom, \textcolor{blue}{ 
thus modifying} the 
chemical and physical properties of the alumina surface. For CH$_4$ 
molecules, a few C atoms bond to Al atoms, while some H atoms bond to 
oxygen and aluminum atoms. Dissociation was observed in all cases, 
with the most likely dissociation channel being CH$_4 \rightarrow $ CH$_2$ 
+ H$_2$, suggesting that this process 
could be used for hydrogen 
production from methane gas, as reported experimentally in the 
literature \cite{YOO201574}. CO$_2$ molecules were observed to 
dissociate in all cases, with the common dissociation channel being 
CO$_2 \rightarrow $ CO + O, producing multiple O atoms upon 
interaction with the surface. The adsorbed C and O atoms were found to 
accumulate around Al atoms, which could be related the adsorption 
mechanism observed in some experiments \cite{LGang}. 
\textcolor{blue}{A video showing the visualization of dissociation
processes is included in the supplementary material.}
\section{Concluding remarks}
\label{sec:conl}

In this work, we present a study of physisorption pathways for $\alpha$--
alumina and several molecules are computed to describe the binding of 
the O, C, or H atoms to the Al and O atoms of the surface 
for prospective studies of chemical 
reactions and applications in catalysis and sensor technologies.
We performed \textcolor{blue}{simulations to calculate} 
the binding energies of $\alpha$--alumina 
with Al atoms layered at the top of the surface named Al--terminated 
and a second one defined as Ox--terminated by O atoms at the top 
surface with O$_2$,  H$_2$O, CH$_4$, 
and CO$_2$ by the self--consistent--
charge density--functional tight--binding (SCC--DFTB) method.
The surfaces are modeled by slabs with thicknesses 
determined by convergence of the optimized
lattice parameters, in good agreement with other experimental
and theoretical results.
We follow several physisorption pathways considering
different molecular orientations where the most stable physisorbed
state on the Al--terminated surface is above the Al atom, while the
Ox--terminated state is found to be above the oxygen atom; the
former is more stable due to the lower physisorption surface energy.
DFTB results show that by comparing the binding energies 
of the adsorption complexes the preferred adsorption sites are 
dependent on the surface structure, Al--terminated and Ox--terminated, 
as well as the orientation of the molecules, as expected. 

Quantum-classical molecular dynamics simulations were performed to investigate the adsorption and dissociation mechanisms. We noted that O$_2$ molecules were mainly adsorbed by the alumina surface, while CH$_4$ molecules were less adsorbed, producing atomic and molecular hydrogen upon interacting with the surface. These molecules also improved the optical adsorption of the alumina surface in the visible range. Based on our calculations, we suggest that the SCC-DFTB approach is suitable for the analysis of dynamical adsorption mechanisms by considering alumina surfaces with an electronic description.

Quantum--Classical Molecular Dynamics simulations are performed 
to investigate the adsorption and dissociation mechanism.
\textcolor{blue}{We noted that} O$_2$ molecules are mainly adsorbed 
by the alumina surface, \textcolor{blue}{while} CH$_4$ molecules are 
less adsorbed, producing atomic and molecular hydrogen \textcolor{blue}{upon} 
interacting with the surface. 
These molecules also improved the optical adsorption of 
the alumina surface in the visible range.
\textcolor{blue}{Based on our calculations}, we suggest that the 
SCC--DFTB approach is suitable for the analysis of dynamical 
adsorption mechanisms by considering alumina surfaces with an 
electronic description.

\medskip
\textbf{Supporting Information} \par 
Supporting Information is available from the Wiley Online Library or from the author.

\medskip
\textbf{Acknowledgements} \par 
We acknowledge support from the European Union Horizon 2020 research
 and innovation program under grant agreement no. 857470 and from the 
 European Regional Development Fund via the Foundation for Polish 
 Science International Research Agenda PLUS program grant 
 No. MAB PLUS/2018/8.
 We acknowledge the computational resources 
 provided by the High Performance Cluster at the National Centre 
 for Nuclear Research in Poland.

\medskip

%


\bibliography{references}
\bibliographystyle{MSP}




\end{document}